\documentclass[12pt,tightenlines,eqsecnum,floats,showpacs,nofootinbib,amsmath,amssymb,aps,prd]{revtex4}

\usepackage{amssymb,amsfonts,amsthm,amscd,amsmath}
\usepackage{graphicx}
\usepackage{enumerate} 
\usepackage{colordvi} 
\usepackage{units}
\usepackage{epsfig}
\usepackage{color}
\usepackage{hyperref}

\hypersetup{
pdfstartview = {FitH},
}

\newcommand{\ket}[1]{|#1\rangle}
\newcommand{\braket}[2]{\langle #1|#2\rangle}
\newcommand{\expval}[3]{\langle #1|#2|#3\rangle}

\newcommand{\abs}[1]{\left|#1\right|}

\DeclareMathOperator{\Tr}{Tr}
\DeclareMathOperator{\R}{Re}
\DeclareMathOperator{\I}{Im}

\begin{document}

\title{A Homogeneous Model of Spinfoam Cosmology}
\author{Julian Rennert$^{1,2}$}
\email{Rennert@stud.uni-heidelberg.de}
\author{David Sloan$^3$}
\email{djs228@hermes.cam.ac.uk}
 \affiliation{$^1$ Centre de Physique Th\'{e}orique 1, CNRS-Luminy, Case 907, F-13288 Marseille\\
$^2$ Institut f\"{u}r Theoretische Physik, Universit\"{a}t Heidelberg, Philosophenweg 16, D-69120 Heidelberg\\
 $^3$ DAMTP, Center for Mathematical Sciences, Cambridge University, Cambridge CB3 0WA, UK}

\begin{abstract}
We examine spinfoam cosmology by use of a simple graph adapted to homogeneous cosmological models. We calculate dynamics in the isotropic limit, and provide the framework for the anisotropic case. We calculate the transition amplitude between holomorphic coherent states on a single node graph and find that the resultant dynamics is peaked on solutions which have no support on the zero volume state, indicating that big bang type singularities are avoided within such models. 
\end{abstract}


\maketitle

\newpage

\section{Introduction}

Cosmological models within the spinfoam framework serve a dual purpose \cite{towards}: Their primary function is to form a proposal for extracting cosmological predictions from a full theory of quantum gravity. These models also perform a useful secondary role in forming a bridge between the canonical \cite{Abhay} and covariant \cite{zakopane} formulations of Loop Quantum Gravity (LQG). The covariant, or spinfoam, approach is a `bottom up' construction - one predicates a quantum model and thence derives dynamics. As such the existence of a semi-classical limit and its agreement with the predictions of General Relativity are not a foregone conclusion, but rather must be examined within physical scenarios. This situation contrasts that of Loop Quantum Cosmology (LQC) \cite{APS}, the application of the principles of LQG to cosmological mini-superspaces. 

Another important question that has to be clarified by spinfoam cosmology is whether physical predictions such as the resolution of cosmological singularities can also be derived within this approach. The resolution of the big bang singularity, and its replacement with a deterministic bounce, is a key success of the canonical theory. The absence of strong singularities is a well trusted result in LQC in the $k=0$ \cite{bojo2, ParamLQCNeverSingular}, and $k=\pm 1$ \cite{ParamFrancesca} FLRW models, and has been extended to include Bianchi I spacetimes \cite{ParamB1Singular}. It forms the basis of investigation of observable consequences of the theory \cite{Measure, AbhayWillIvan}. It is therefore a crucial test of the spinfoam approach that it reproduces these features in the ultraviolet sector. 

In \cite{towards} it was shown how to calculate the transition amplitude between two quantum states of gravity in the homogeneous and isotropic cosmological regime using a simple two-node graph (the dipole graph) at the first order in the vertex expansion. The main result of this work was to demonstrate that the used new spinfoam vertex amplitude (EPRL/KKL), \cite{spinfoam1, spinfoam2, spinfoam3, spinfoam4}, together with some other ingredients, are adequate to derive a classical limit which can be identified with the Friedmann dynamics of an empty flat, homogeneous and isotropic universe, i.e. (static) Minkowski space. This result was further strengthened in \cite{cosmoconstant}, where a slight modification of the spinfoam vertex was utilized to implement a cosmological constant and derive a de Sitter universe as the classical limit. However, despite these original results being interesting, they exhibit a considerable deficiency, namely they fail to reproduce the curvature term $\frac{k}{a^2}$, which appears in the Friedmann equation. Such a term is expected, since the chosen graph is dual to a (degenerate) triangulation of the three sphere, (the closed topology, in which $k=1$). The authors of \cite{towards} argue that this term might be recovered by taking higher orders of the spin approximation into account. However, this term appears in a more natural manner, as we will show in section \ref{ClassicalPrelim}. 

Our approach is the following: We will examine flat ($k=0$) Friedmann-Lem\^aitre-Robertson-Walker (FLRW) models by use of a simple graph. This `Daisy' graph consists of a single node which is both the source and target of three links. This graph can be thought of in two equivalent ways: In the first instance one has tessellated space by identical cubes, and so by symmetry opposite faces of a cube are identified, thus an outgoing edge dual to a given face is an incoming edge dual to the opposite face. The second instance is to consider the spatial slice to be a flat three-torus, with each link transcribing a compact direction. This is what separates this approach from the cubulation used in \cite{cubulations} and also in \cite{quantred}.

The second motivation is to provide the framework for investigating the more complicated case of Bianchi I cosmologies. Since the inclusion of matter within the spinfoam paradigm has not yet been fully realized, one can only investigate FLRW models with trivial classical dynamics. In the anisotropic homogeneous systems comprising the Bianchi models there is a rich physical evolution even in the absence of any matter. These models have been examined extensively in LQC, both within the quantum framework \cite{EdB1,EdB2,EdB9} and the semi-classical effective framework \cite{AlexB1,DahWei,Cailleteau,DahWei2,Maartens,Gupt,AlexB2,Gupt2,AlexB9} and it would be a strong evidence for the validity of the spinfoam cosmology approach if one could derive these models from the full quantum theory.

\section{Recap of the theoretical framework}

Let us briefly review the necessary theoretical input to make our ideas and calculations tractable, and fix our notation. Since we rely heavily on the theory as introduced in \cite{towards} we refer the reader to the original source or \cite{rovelli, zakopane} for a more detailed discussion.

\subsection{LQG and spinfoams}

The kinematical Hilbert space of LQG is defined as the direct sum of subspaces $\mathcal{H}_{\Gamma}$ over all graphs $\Gamma$, embedded in a three dimensional manifold $\Sigma$. Since we want to work in a cosmological regime, describing just a finite number of degrees of freedom, it is sufficient for us to consider just one of these subspaces. This Hilbert space $\mathcal{H}_{\Gamma}$ is defined on a graph $\Gamma$ with $L$ links and $N$ nodes. Its elements are the spin network functions; gauge invariant, square integrable functions $\Psi : SU(2)^L \rightarrow \mathbb{C}$, (\textit{holonomy representation}). Since gauge transformations act on the nodes N the Hilbert space $\mathcal{H}_{\Gamma}$ is given by
\begin{equation}
\mathcal{H}_{\Gamma} = \mathcal{L}^2(SU(2)^L/SU(2)^N)\,.
\label{eq:1}
\end{equation}

The name holonomy representation results from the circumstance that the $SU(2)$ elements $h_l$ are the holonomy of the Ashtekar-Barbero connection along the link $l$, i.e.
\begin{equation}
h_l = h_l(A) = \mathcal{P} \exp \left(\int_l{A}\right)
\label{eq:2}
\end{equation}

with $A = A^i_a \tau^i \text{d}x^a$. The components of $A$ are given by $A^i_a = \Gamma^i_a + \gamma K^i_a$ with $\Gamma^i_a$ being the spin-connection, $K^i_a$ the extrinsic curvature of $\Sigma$ and $\gamma \in \mathbb{R}_{>0}$ is the Barbero-Immirzi parameter. Thus the $SU(2)$ elements $h_l$ contain the geometrical information of the quantum state $\Psi(h_l)$.

Another representation, related to the former one via the Peter-Weyl transformation \cite{zakopane, holo}, is the spin-intertwiner representation. In this representation the graph $\Gamma$ carries spins $j_l \in \frac{\mathbb{N}}{2}$ at each link and invariant tensors $i_n$, called intertwiners, at each node. Those spins correspond to the spins of the unitary irreducible representations of $SU(2)$ and the intertwiners belong to the $SU(2)$-invariant subspace $\mathcal{K}_n = \text{Inv}_{SU(2)} [\mathcal{H}_n]$, where $\mathcal{H}_n$ is the tensor product of the representation spaces carried by the links meeting at the node $n$, $\mathcal{H}_n=\otimes_{l \in n} \mathcal{H}_{j_l}$. A general state in $\mathcal{H}_{\Gamma}$ has the following structure in the spin-intertwiner representation
\begin{equation}
\Psi_{j_l,i_n}(h_l) = \left(\bigotimes_n i_n\right) \cdot \left(\bigotimes_l \mathcal{D}^{(j_l)}(h_l)\right)\,,
\label{eq:3}
\end{equation}

where $\mathcal{D}^{(j_l)}(h_l)$ is the $2j_l + 1$ dimensional Wigner matrix of the holonomy $h_l$ and the dot indicates contraction of indices.

There are now two interpretations of the 3-dim. manifold $\Sigma$ in which $\Gamma$ lives. First one can imagine $\Sigma$ to be a spacelike slice at some coordinate time $t$. The spinfoam model would then define an amplitude from $\mathcal{H}_{\Gamma(\Sigma_t)}$ to $\mathcal{H}_{\Gamma(\Sigma_{t+1})}$ which allows us to interpret this amplitude as a transition amplitude between two states of geometry on the spatial slice. The second interpretation holds $\Sigma$ to be a 3-dim. boundary of a 4-dim. spacetime region. The states in $\mathcal{H}_{\Gamma(\Sigma)}$  are thus not thought of as `states at some time', but rather as \textit{boundary states}, \cite{oeckl1, oeckl2, oeckl3}. We note that the first case is a special case of the second one for disconnected spatial boundaries.

The dynamics of these quantum states can be defined via the spinfoam formalism. Think again of a boundary state $\Psi \in \mathcal{H}_{\Gamma}$ with $\Gamma \subset \Sigma$. A spinfoam lives on a 2-complex made up of vertices, edges and faces. A 2-complex can be seen as a discretization of 4-dim. spacetime and heuristically may be thought of as resulting from a canonical spin network evolving in time. Even if one deals with the diffeomorphism invariant s-knot states one has to consider an explicit embedding in order to calculate holonomies and fluxes for a given (patch of) spacetime. Thus, we work with embedded graphs to facilitate contact with the canonical formulation (i.e. LQC) in which most work to date has been performed. This embedding enables the direct projection of established holonomies and fluxes representing the FRW geometry onto our network. Overall, this picture leads us to view also our spinfoam to be embedded in spacetime which contrasts the viewpoint of abstract non-embedded spinfoams. Either way a spinfoam model assigns an amplitude to the state $\Psi$ in the following way
\begin{equation}
\braket{W}{\Psi} = \int{dh_l \, W(h_l) \Psi(h_l)}\,,
\label{eq:4}
\end{equation}

where $W(h_l)$ is given by the EPRL/FK spinfoam model \cite{spinfoam1, spinfoam2, spinfoam3, spinfoam4, holo} and is given by
\begin{equation}
W(h_l) = \sum_{\sigma}{\int{dh^{\text{bulk}}_{vl} \prod_{f \subset \sigma}{\delta(h_f)} \prod_{v \subset \sigma}{A_v(h_{vl})} }}\,.
\label{eq:4.2}
\end{equation}

The sum ranges over spin network histories, the $h_f$ are the holonomies around a face and $A_v(h_{vl})$ is called the vertex amplitude. We will present its precise structure in section \ref{ourmodel} where we again will follow closely \cite{towards}. As explained above, for a disconnected boundary $\Sigma$ this amplitude is interpreted as a transition amplitude and captures the probability for a state on $\Sigma_{t_1}$ to evolve into another (or the same) state on $\Sigma_{t_2 > t_1}$. Thus, we are not going to calculate the explicit time evolution of states nor the full time evolution operator but probability amplitudes for single coherent states.

\subsection{Coherent states}

Coherent states are an important tool for the examination of the classical limit of any quantum theory. In this section we will summarize a few definitions about the coherent states for LQG. In particular we will use the Livine-Speziale coherent intertwiners \cite{newvertex} as well as the coherent states in the holomorphic representation \cite{coherent, holo} later in this work. 

The Livine-Speziale coherent intertwiners make use of the Perelomov coherent states for $SU(2)$ such that the intertwiner $i_n$ in (\ref{eq:3}) is replaced by a coherent intertwiner. A Perelomov coherent state for $SU(2)$ takes the highest weight state $\ket{j,j} \in \mathcal{H}^{(j)}$, which is a coherent state along $\hat{e}_z$, and rotates it with a Wigner matrix $\mathcal{D}^{(j)}(h_{\vec{n}})$ such that it is coherent along another axis $\vec{n}$. The element $h_{\vec{n}} \in SU(2)$ corresponds to the $SO(3)$ element $R_{\vec{n}}$ that rotates $\hat{e}_z$ into $\vec{n}$. Thus we obtain the coherent state $\ket{j,\vec{n}} \equiv \mathcal{D}^{(j)}(h_{\vec{n}}) \ket{j,j}$. Consider a node $n$ which joins $E$ links $e$ together. A coherent intertwiner at this node $n$ is now given by the tensor product of the coherent states coming from each single link. The gauge invariance of these states is achieved via group integration.
\begin{equation}
\Phi_n(\vec{n}_e) = \int_{SU(2)}{\text{d}g \bigotimes^E_{e=1} \mathcal{D}^{(j_e)}(g) \ket{j_e,\vec{n}_e}}
\label{eq:5}
\end{equation}

The holomorphic coherent states are characterized by an element $H_l \in SL(2,\mathbb{C})$ given at each link of the graph $\Gamma$. They are defined by
\begin{equation}
\Psi_{H_l}(h_l) = \int_{SU(2)^N}{\text{d}g_n \prod_l{K_t(g_{s(l)} h_l g_{t(l)}^{-1}, H_l)}}\,,
\label{eq:6}
\end{equation}

where $K_t$ is the analytic continuation of the $SU(2)$ heat kernel to $SL(2,\mathbb{C})$ and the group integration again ensures gauge invariance. The heat kernel is given by
\begin{equation}
K_t(a,B) = \sum_{j \in \mathbb{N}_0/2}{(2j+1) \, e^{-\alpha t j (j+1)} \, \text{Tr}(\mathcal{D}^{(j)}(a B^{-1}))}
\label{eq:7}
\end{equation}

with $a \in SU(2)$, $B \in SL(2,\mathbb{C})$ and $\alpha, t \in \mathbb{R}_{>0}$. The $SL(2,\mathbb{C})$ label $H_l$ now allows for two different decompositions \cite{holo}. The first one is the polar decomposition 
\begin{equation}
H_l = h_l(A) \, \exp \left(i \frac{E_l}{8 \pi G \hbar \gamma} t_l\right)
\label{eq:8}
\end{equation}

and shows clearly that $H_l$ determines a point in classical phase space on which the coherent state is peaked. $h_l \in SU(2)$ is the holonomy of the Ashtekar connection $A^i_a$ and $E_l \in \mathfrak{su}(2)$ is the flux of the densitized triad $E^a_i$. Thus, a coherent state with label (\ref{eq:8}) corresponds to a classical configuration $(A^i_a,E^a_i)$.

The second decomposition of $H_l$ uses two $SU(2)$ elements $h_{\vec{n}_l}$ and $h_{\vec{n}^{\prime}_l}$ which, analogously to the $SU(2)$ elements of the Perelomov coherent states, correspond to the transformation of $\hat{e}_z$ into $\vec{n}_l$ and $\vec{n}^{\prime}_l$. Furthermore, a complex number $z_l$ is used whose real part is associated to the extrinsic curvature and its imaginary part is related to the area that is pierced by the link $l$ \cite{holo}.
\begin{equation}
H_l = h_{\vec{n}_l} \, e^{-i z_l \frac{\sigma^3}{2}} \, h^{-1}_{-\vec{n}^{\prime}_l}
\label{eq:9}
\end{equation}

We denote the real and the imaginary part of $z_l$ as $z_l=c_l + i p_l$ and $\sigma^3$ is the third Pauli matrix. The relation between the two decompositions becomes clear by writing (\ref{eq:9}) in the polar decomposition. One finds that \cite{holo}
\begin{equation}
h_l = h_{\vec{n}_l} \, e^{-i c_l \frac{\sigma^3}{2}} \, h^{-1}_{-\vec{n}^{\prime}_l}\,,
\label{eq:10}
\end{equation}

\begin{equation}
E_l = \int_{f_l}{E} = 8 \pi G \hbar \gamma \, p_l \, \vec{n}^{\prime}_l \cdot \frac{i \vec{\sigma}}{2 t_l}\,.
\label{eq:11}
\end{equation}

Where $f_l$ is the face dual to the link $l$ with area $\mathcal{A}_l = 8 \pi G \hbar \gamma \, p_l / t_l$. Other coherent states, based on the so called flux representation for LQG were introduced in \cite{OPS1, OPS2}. These states posses a slighly different peakedness behaviour for the mean value of the flux operator which derives from a modified heat equation on $SU(2)$ using a different Laplacian, \cite{OPS3}. For future research it might be interesting to consider these states instead of the above presented ones. However, in order to be able to compare our results with \cite{towards} we stick to the holomorphic coherent states in this work.

\subsection{Classical preliminaries}
\label{ClassicalPrelim}

We are interested in the applicability of spinfoam cosmology to homogeneous models, both in the isotropic and anisotropic cases. In this section we will establish the holonomies and the fluxes for such models. We assume our spacetime to be of the form $\mathcal{M} = \mathbb{R} \times \Sigma$, with $\Sigma$ being a homogeneous $3$-space. Under the additional assumption of isotropy the metric of $\mathcal{M}$ can be given by
\begin{equation}
ds^2 = -dt^2 + a(t)^2 d \Omega^2
\label{eqcl:1}
\end{equation}

with $d \Omega^2 = dr^2 / (1-kr^2) + r^2 d\theta^2 + r^2 \sin^2\theta \, d\phi^2$ and $k \in \{0,\pm1\}$. The parameter $k$ distinguishes three different spaces with constant curvature, where we are interested in the closed $(k=1)$ and the flat $(k=0)$ case. The flat and closed universes are special cases of the Bianchi I and IX universes respectively, in which all scale factors have been identified. If we consider a universe without matter but just a cosmological constant $\Lambda$, the metric (\ref{eqcl:1}) evolution obeys the Friedmann equation
\begin{equation}
\left(\frac{\dot{a}}{a}\right)^2 + \frac{k}{a^2} = \frac{\Lambda}{3}\,.
\label{eqcl:2}
\end{equation}

In the case of vanishing cosmological constant the only possible solution is a static spacetime $a(t) = const.$, where for $k=0$ one recovers Minkowski space. If $\Lambda \neq 0$ one obtains for $k=0$, and under the assumption that $a(t) \gg 1$ also for $k=1$, the \textit{de Sitter solution}, $a(t) = \exp(\pm \sqrt{\Lambda/3} \, t)$.

If we drop the restriction to isotropic models we obtain a Bianchi I universe in the flat case, which is described by the following line element
\begin{equation}
ds^2 = -dt^2 + a_1(t)^2 dx^2 + a_2(t)^2 dy^2 + a_3(t)^2 dz^2\,.
\label{eqcl:3}
\end{equation}

Considering again a vacuum spacetime (with $\Lambda=0$), the three directional scale factors $a_1, a_2, a_3$ have to satisfy
\begin{equation}
a_1 \dot{a}_2 \dot{a}_3 + a_2 \dot{a}_1 \dot{a}_3 + a_3 \dot{a}_1 \dot{a}_2 = 0\,.
\label{eqcl:4}
\end{equation}

This equation is solved by the so called \textit{Kasner universe} and is given by $a_i(t) = t^{\kappa_i}$. The Kasner exponents have to fulfill the conditions $\sum_i{\kappa^2_i} = \sum_i{\kappa_i} = 1$. From those conditions one deduces that one exponent has to be negative, while the other two are positive which leads to a contraction in one direction and an expansion in the other two (the standard choice is $\kappa_1 = -1/3$ and $\kappa_2 = \kappa_3 = 2/3$).

Now, in order to specify the holonomy and the flux, we need the Ashtekar connection and the corresponding densitized triad. For that we will use the results provided in \cite{bojo4, bojo3, bojo2, bojo1}.

In a general, i.e. non-cosmological, setting the Ashtekar connection is given by $A^i_a = \Gamma^i_a + \gamma K^i_a$, with $K^i_a$ beeing related to the extrinsic curvature 
\begin{equation}
K^i_a = e^{ib} K_{ab} = \frac{1}{2} e^{ib} \mathcal{L}_{(\frac{\partial}{\partial t})} h_{ab}\,,
\label{eqcl:5}
\end{equation}

where the $e^i_a$ are co-triads, such that the spatial metric can be expressed as $h_{ab} = \delta_{ij} e^i_a e^j_b$. The connection coefficients $\Gamma^i_a$ are calculated via contraction of the spin connection $\Gamma^i_a = -\frac{1}{2} \varepsilon^{ijk} \theta_{ajk}$, which is given by
\begin{equation}
\theta^{\,\,i}_{a\,j} = -e^b_j \left(\partial_a e^i_b - \Gamma^c_{ab} e^i_c\right)\,.
\label{eqcl:6}
\end{equation}

$\Gamma^c_{ab}$ is the Levi-Civita connection compatible with $h_{ab}$, expressed in terms of the co-triads. However, using the framework of invariant connections on principal fibre bundles, as explained in \cite{bojo4}, simplifies the tedious calculation of $A^i_a$ via (\ref{eqcl:5}) and (\ref{eqcl:6}) enormously. Now, a Bianchi model is a symmetry reduced model of general relativity by a symmetry group $S$, which acts freely and transitively on $\Sigma$. If $\Sigma$ is invariant under the action of $S$ it is an homogeneous 3-space and a connection can be decomposed as $A^i_a = \phi^i_I \omega^I_a$, with left invariant 1-forms $\omega^I_a$ and constant coefficients $\phi^i_I$. A further reduction, which leaves us for example with the three gauge invariant degrees of freedom in (\ref{eqcl:4}), is achieved by diagonalizing $\phi^i_I$. This has the effect that we can write
\begin{equation}
A^i_a = c_{(K)} \Lambda^i_K \tilde{\omega}^K_a\,,
\label{eqcl:7}
\end{equation}

with $\Lambda^i_K \in SO(3)$, \cite{bojo2}. Using the left-invariant vector fields $X^a_I$, dual to the 1-forms $\omega^I_a$, allows us to decompose also the densitized triad as
\begin{equation}
E^a_i = p^I_i X^a_I = p^{(K)} \Lambda^K_i \tilde{X}^a_K\,,
\label{eqcl:8}
\end{equation}

where the second equality results again from a diagonalization of $p^I_i$. These six coefficients $(c_K, p^K)$, $K=1,2,3$ now span the phase space of our reduced homogeneous model with the symplectic structure \cite{EdB1}
\begin{equation}
\{c_I,p^J\} = \frac{8 \pi G}{3} \gamma \delta^J_I\,.
\label{eqcl:9}
\end{equation}

If we expand the co-triads as $e^i_a = e_{(K)} \Lambda^i_K \tilde{\omega}^K_a$, with arbitrary $e_K \in \mathbb{R}$, we get the following relations (no summation)
\begin{equation}
p^I = \left|\varepsilon_{IJK} \, e_J e_K\right| \text{sgn}(e_I)\,.
\label{eqcl:10}
\end{equation}

With these simplifications the connection components $\Gamma^i_a = \Gamma_{(I)} \Lambda^i_I \tilde{\omega}^I_a$ are given by, (no summation, even permutation of \{1,2,3\})
\begin{equation}
\Gamma_I = \frac{1}{2} \left(\frac{p^K}{p^J} n^J + \frac{p^J}{p^K} n^K - \frac{p^J p^K}{(p^I)^2} n^I\right)\,.
\label{eqcl:11}
\end{equation}

The $n^I$ characterize our Bianchi model, we have for example $n^I = 0$ for Bianchi I and $n^I = 1$ for Bianchi IX. Thus, we see that the Bianchi I models have vanishing spin connection $\Gamma^i_a$. The extrinsic curvature is given by $K_I = \frac{1}{2} \dot{e}_I$, \cite{bojo2}, where the dot indicates a derivative with respect to the coordinate time $t$. Now, we find the following results for the Ashtekar connection in the homogeneous setting
\begin{equation}
A^i_a = c_{(I)} \Lambda^i_I \tilde{\omega}^I_a = \left(\Gamma_{(I)} + \frac{\gamma}{2} \dot{e}_{(I)}\right) \Lambda^i_I \tilde{\omega}^I_a\,.
\label{eqcl:12}
\end{equation}

In the isotropic case we have $p^1 = p^2 = p^3$ and (\ref{eqcl:11}) gives us $\Gamma_I = 0$ in the flat case (Bianchi I), whereas we get $\Gamma_I = \frac{1}{2}$ in the model with positive curvature (Bianchi IX). We can thus write $c = \frac{1}{2}(k + \gamma \dot{e})$, $k \in \{0,1\}$. For the anisotropic (Bianchi I) model we get $c_I = \frac{\gamma}{2} \dot{e}_I$.

Before we apply this formalism to our one-node graph let us make the following observation. In \cite{towards} it was shown that the holomorphic transition amplitude between two homogeneous and isotropic quantum states, which are supposed to correspond to a curved geometry $(k=1)$, is given by
\begin{equation}
W(z) = N \, z \, \exp \left(- \frac{z^2}{2 t \hbar}\right)\,.
\label{eqcl:13}
\end{equation}

Following the reasoning in \cite{cosmoconstant} the main contribution of $W(z)$ is obtained when the real part of $z^2$ vanishes and its imaginary part is proportional to $\pi l$, $l \in \mathbb{Z}$. Now, if we use the correct relation (which was already noted in \cite{triangulated}) between $c$ and the metric variables, i.e. $c=\R (z) = \frac{1}{2} (k+\gamma \dot{a})$, instead of just $c= \gamma \dot{a}$ we can reproduce the correct Hamiltonian constraint. Therefore, we require that the real and the imaginary part (which doesn't contribute anyway, if we consider $\abs{W(z)}$) vanish. Thus, we get from $z^2 = (c+ip)^2$
\begin{equation}
c^2 - p^2 \stackrel{!}{=} 0\,.
\label{eqcl:14}
\end{equation}

However, the $p^2$ term will disappear if we consider the proper normalized amplitude as done in \cite{cosmoconstant} or \cite{thiemann3}. Thus, we find $c^2 = 0$ and
\begin{align}
c^2 &= \frac{1}{2} \left(\frac{1}{2} + \gamma \dot{a} + \frac{\gamma^2 \dot{a}^2}{2}\right) = 0\notag\\
&= \frac{1}{2} (1+\gamma \dot{a}) - \frac{1}{4} \left(1-\gamma^2 \dot{a}^2\right)\notag\\
&= c - \frac{1}{4} \left(1-\gamma^2 \dot{a}^2\right)
\label{eqcl:15}
\end{align}

\begin{equation}
\Rightarrow \quad - \frac{1}{4} \left(1-\gamma^2 \dot{a}^2\right) = 0
\label{eqcl:16}
\end{equation}

Scaling of $\dot{a}$ and multiplication by $a$ gives us
\begin{equation}
- \dot{a}^2 a + a = 0
\label{eqcl:17}
\end{equation}

which is the correct Hamiltonian constraint for a curved FLRW universe \cite{bojo5}. In this paper we are interested in flat spatial slices but it is imaginable to include curvature analogously in our model (i.e. using the correct Ashtekar connection). However, one should note that the graph structure must also support the topology under consideration and thus one might be forced to choose a different graph to probe a curved spacetime, e.g. as done in \cite{cubulations, quantred}.

We have already mentioned the definition of the holonomy in (\ref{eq:2}). Now, let us define the flux. If we denote the link along which we evaluate the holonomy by $l$ then $S_l$ denotes a surface pierced by $l$. One says $S$ is dual to $l$. The flux of the electric field $E=E^a_i \tau^i X_a$ through a surface $S_l$ is given by
\begin{equation}
E(S) = \int_{S_l}{(\ast E)^j n^j}\,,
\label{eqcl:18}
\end{equation}

where $\ast$ denotes the Hodge dual, which converts our vector $E$ into a 2-form, ($\text{dim}(\Sigma)=3$), and $n^j = n^i \tau^i$ is a $\mathfrak{su}(2)$ valued scalar smearing function \cite{thiemann}. We will use the definition
\begin{equation}
(\ast E) = (\ast E)^j \tau^j = (\ast E)^j_{a_1 a_2} \text{d}x^{a_1} \wedge \text{d}x^{a_2} \tau^j
\label{eqcl:19}
\end{equation}

with
\begin{equation}
(\ast E)^j_{a_1 a_2} = \varepsilon_{a a_1 a_2} E^a_j\,.
\label{eqcl:20}
\end{equation}

These definitions will become necessary especially for the anisotropic case, when we explicitly have to calculate the Ashtekar connection and the flux for our model.

\section{Our Model}
\label{ourmodel}

In this section we want to calculate the transition amplitude between two flat, homogeneous and isotropic universes using the spinfoam formalism. As is customary in (quantum) cosmology, we are interested in the largest wavelength modes and ignore shorter scale fluctuations. Our model can be interpreted in two ways: Either as probing the universe on the largest scales, in which only the largest wavelength is relevant, or equivalently as tessellating space with cubes and restricting the geometry to homogeneity thereupon. Ideally one should take a large number of such cubes and consider all fluctuations away from homogeneity in the calculation of transition amplitudes and then coarse-grain for large scale behaviour. However, in practice this is highly impractical and therefore we follow the usual philosophy applied in cosmology and symmetry reduce before establishing dynamics. Despite the inherent shortcomings of such a simplification, this has proven highly effective in classical cosmological approaches, and is the basis of all quantum cosmologies.

\begin{figure}[ht]
\includegraphics[width=0.4\textwidth]{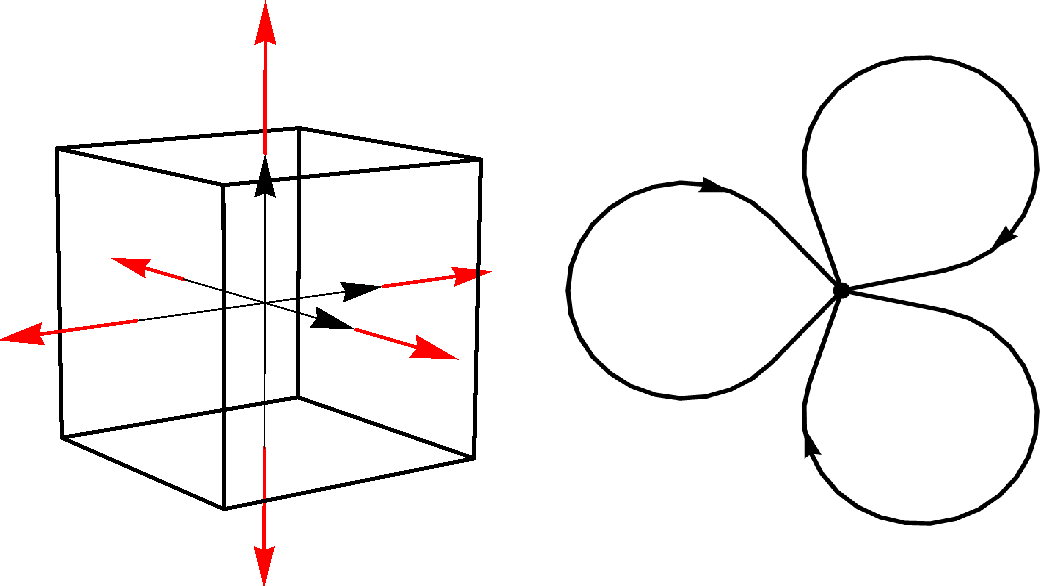}
\caption{Cube and Daisy graph}
\label{fig1}
\end{figure}

In the spinfoam cosmology approach this means that on the one hand we have to identify certain homogeneous states which are presumably characterized by a certain subclass of all possible graphs and a certain (homogeneous) coloring. On the other hand, given that the spinfoam formalism is considered as a non-perturbative framework for quantum gravity, we have to employ a truncation of the full quantum dynamics. Therefore, we think of a cubical partition of 3-space $\Sigma$. Homogeneity then allows us to restrict our considerations to a single cube whose dual graph (with toroidal topology) is given by the \textit{Daisy graph}\footnote{Note that this graph is not necessarily equivalent to that employed in \cite{towards}, since we aim for a topologically distinct scenario. A key aim of this paper is to address flat ($k=0$) cosmologies whereas the dipole graph is supposed to describe a closed ($k=1$) geometry.}, see FIG.(\ref{fig1}). It is not hard to see that the restriction to a smaller graph corresponds to a truncation of degrees of freedom at the kinematical level. But it is also true that this does not automatically imply a cosmological setting. In fact, one can certainly build homogeneous states by using a larger lattice and keeping all holonomies and fluxes the same. Homogeneity then allows us to identify all lattice points, and thus the simplification made is appropriate. Note further, that the identification with cosmology also arises because of the particular holonomies and fluxes which we are using: In this sense we identify our simple graph with a cosmological setting. The original motivation for this graph, especially the use of three closed links, was its potential applicability to anisotropic cosmological settings and therewith a physically more complex situation, a problem we will tackle in a follow up paper \cite{ournextpaper}. In this paper we will restrict our attention mostly to the isotropic case and see that this one node graph is already sufficient to reproduce the original result of \cite{towards}.

Let us furthermore point out that, unless one is dealing with a symmetry reduced dynamics, the regime in which a graph provides the basis for a good homogeneous state, may depend on the full quantum dynamics. This means that by allowing for larger quantum fluctuations, i.e. a more complicated dynamics, two different graphs, which were originally thought to describe the same homogenous state, may lead to different results\footnote{We are indebted to an anonymous referee for pointing this out to us.}. This closely relates to our use of the one-vertex spinfoam expansion and the objective of finding an effective dynamics from the full, non-perturbative quantum dynamics as pursued for example in \cite{gft}. In spinfoam cosmology calculations to date were all performed using a single spinfoam vertex to specify the dynamics. The rationale behind this approxmation, next to its calculability, is beautifully elaborated on in \cite{sfexp1} and \cite{sfexp2}. There it is shown, in a simple discretized parametrized model, how an expansion in a small number of vertices allows one to achieve good agreement with the continuum model, both in the classical and the quantum regime. Note that approximating the number of spinfoam vertices is not to be confused with a semiclassical approximation in terms of a dimensionful parameter such as $\hbar$. Note furthermore, that in TQFT the result of transition amplitudes does not depend on the underlying triangulation. Now quantum gravity is certainly not a topological theory, however, the point is, that in certain regimes it may behave similarly and thus is not sensitive whether one uses a finer or more complicated bulk triangulation (`Ditt-invariance'). Of course, the goal has to be to gradually increase the number of spinfoam vertices. 

Another interesting approach towards the extraction of a cosmological scenario was recently obtained within the Group Field Theory approach to quantum gravity \cite{gft}. The GFT approach offers some promising features concerning the identification of general homogeneous states, independent of the underlying graph structure, and the mentioned interplay with the full quantum dynamics. This may allow for an inclusion of inhomogeneities and may also provide a possibility to calculate corrections to the Friedmann equation. Eventually, one would like to compare predictions coming from both models.

\subsection{The Setting}
First, let us recall the definition for the vertex amplitude to specify the dynamics. Despite it being shown in \cite{hellmann, lewand} that there exist additional 2-complexes which contribute at the one vertex level (for the dipole graph) we will consider just the single spinfoam history which corresponds to our boundary graph in the sum in (\ref{eq:4.2}). We consider the spinfoam that simply connects the two graph vertices with a single spinfoam vertex. The one vertex spinfoam expansion $(v=1)$ leads to the factorization of our amplitude $\braket{W}{\Psi}\, $\footnote{In \cite{hellmann} this factorization of the amplitude was critisized because it indicates that there is no dynamics or evolution of degrees of freedom and we agree that in future work this problem has to be treated with more care. However, for the moment we go along with this strategy mainly for its computability but also for the following reasons: First, in \cite{cosmoconstant} it was shown that using the factorizing amplitude it is possible to reproduce the de Sitter solution, which is a non-static universe. Second, in the follow up paper to this one \cite{ournextpaper} we will derive a anisotropic Bianchi I universe, which too possesses classicaly only dynamical solutions, using the factorizing amplitude. Thus, we leave the examination of non-factorizing transition amplitudes for future research. Let us just mention, that it is not necessary to use higher orders in the vertex expansion but to take the face amplitudes properly into account to avoid factorizing amplitudes.}. The face amplitude $\delta(h_f)$ in Eq.(\ref{eq:4.2}) peaks the $h_{vl}$ onto the $h_l$ and the coherent states $\Psi_{H_l}(h_l)$ are peaked on the $H_l$. With these simplifications and following \cite{towards} the transition amplitude between an initial and a final geometry is given by
\begin{equation}
W(\Psi_f,\Psi_i) = \overline{A_v(H_l(z_f))} \, A_v(H_l(z_i))\,,
\label{eqs:1}
\end{equation}

where the vertex amplitude is given by \cite{holo}
\begin{equation}
A_v(H_l(z)) = \int_{G^{N-1}}{\text{d}G'_n \, \prod^3_{l=1}{\sum_{j \in \mathbb{N}_0/2}{(2j+1) \, e^{-\frac{t}{2} j (j+1)} \, \text{Tr}_j(H_l(z) Y^{\dagger} G_{s(l)} G^{-1}_{t(l)} Y)}}}\,,
\label{eqs:2}
\end{equation}

where $G$ is $SO(4)$ for the Euclidean theory and $SL(2,\mathbb{C})$ for the Lorentzian theory, respectively. As was explained in \cite{reg} for the Lorentzian case we will neglect one integration so that $W(z)$ does not diverge. Furthermore, since our graph has just one node we find that source and target node of each link are the same, i.e. $s(l) = t(l)$, thus leading to $G_{s(l)} G^{-1}_{t(l)} = \mathbb{I}$. 

A clear advantage of using this graph is its simple application in the homogeneous case. It allows us to explicitly calculate the $SL(2,\mathbb{C})$ elements for our coherent states and with that provides helpful insights also for more complicated structures. We begin by calculating the $SL(2,\mathbb{C})$ elements $H_l(z)$ using the decomposition (\ref{eq:9}) 
\begin{equation}
H_l(z) = u_l \, e^{-i z \frac{\sigma^3}{2}} \, \tilde{u}^{-1}_l\,,
\label{eqs:3}
\end{equation}

where $u_l$ and $\tilde{u}_l$ are elements of $SU(2)$. We have three links, $l_1$, $l_2$, $l_3$ and six normal vectors $n_1 = \hat{e}_x$ and $\tilde{n}_1 = -\hat{e}_x$, $n_2 = \hat{e}_y$ and $\tilde{n}_2 = -\hat{e}_y$ and $n_3 = \hat{e}_z$ and $\tilde{n}_3 = -\hat{e}_z$. The normal vectors $n_l$ and $\tilde{n}_l$ are obtained via a $SO(3)$ transformation of $\hat{e}_z$ and the $SU(2)$ elements $u_l$ and $\tilde{u}_l$ are related to these $SO(3)$ transformations, cf. appendix A. Now, we have to bring the three $SL(2,\mathbb{C})$ elements in the following form
\begin{equation}
H_l(z) = e^{-i \alpha_1 \frac{\sigma^3}{2}} e^{-i \beta \frac{\sigma^2}{2}} e^{-i \alpha_2 \frac{\sigma^3}{2}}\,.
\label{eqs:20}
\end{equation}

This means that we have to find the angles $\alpha_1$, $\alpha_2$ and $\beta$\,. Given the $SL(2,\mathbb{C})$ elements $H_l(z)$ in this form we are then able to represent their Wigner matrices for all $j$ if we recall that the angular momentum operators $\hat{J}_x, \hat{J}_y, \hat{J}_z$ are given by $\hat{J}_x = \frac{\sigma^1}{2}$, $\hat{J}_y = \frac{\sigma^2}{2}$ and $\hat{J}_z = \frac{\sigma^3}{2}$ in the $j = \frac{1}{2}$ representation. Hence, we get
\begin{equation}
\mathcal{D}^{(j)}(H_l(z)) = e^{-i \alpha_1 \hat{J}^{(j)}_z} e^{-i \beta \hat{J}^{(j)}_y} e^{-i \alpha_2 \hat{J}^{(j)}_z}\,.
\label{eqs:21}
\end{equation}

One finds the following angles (cf. equation (\ref{eqa:17}), (\ref{eqa:18}), (\ref{eqa:19}) in the appendix)
\begin{align}
H_1(z) &: \quad \alpha_1 = \alpha_2 = \frac{\pi}{2} \quad , \quad \beta = \pi - z\label{eqs:22}\\
H_2(z) &: \quad \alpha_1 = \pi \quad , \quad \alpha_2 = 0 \quad , \quad \beta = \pi - z\label{eqs:23}\\
H_3(z) &: \quad \alpha_1 = z \quad , \quad \alpha_2 = 0 \quad , \quad \beta = 0 \label{eqs:24}
\end{align}

We can now calculate the transition amplitude
\begin{equation}
W(z) = \int_G{dG \: \prod^{3}_{l=1}{\sum_j{d_j \, e^{-2 t \hbar j (j+1)} \Tr\left(\mathcal{D}^{(j)}(H_l(z)) \tilde{G}\right)}}}\,,
\label{eqs:25}
\end{equation}

where we have defined $d_j=2j+1$ and $\tilde{G} \equiv Y^{\dagger} \mathcal{D}^{(j^+,j^-)}(G_s G^{-1}_t) Y$ in the case of Euclidean gravity, $(G \in SO(4))$, or $\tilde{G} \equiv Y^{\dagger} \mathcal{D}^{(\gamma j,j)}(G_s G^{-1}_t) Y$ in the Lorentzian case, $(G \in SL(2,\mathbb{C}))$. For detail cf. \cite{zakopane}. (Despite $G G^{-1} = \mathbb{I}$, because $s(l) = t(l)$ as mentioned earlier , we keep $\tilde{G}$ for completeness.) Lets start by calculating the trace for $l=1$
\begin{align}
&\Tr\left(\mathcal{D}^{(j)}(H_1(z)) \tilde{G}\right) = \sum^{j}_{m=-j}{\expval{j,m}{\, \mathcal{D}^{(j)}(H_1(z)) \tilde{G} \,}{j,m}}\notag\\
&= \sum^{j}_{m,k=-j}{e^{-i (m+k) \frac{\pi}{2}} \: d^{(j)}_{mk}(\pi -z) \: \expval{j,k}{\, \tilde{G} \,}{j,m}}\,,
\label{eqs:26}
\end{align}

where we have inserted a unit operator. This leads to a separation of the geometrical information, stored in $H_l(z)$, and the gauge invariant contribution, given by $\expval{j,k}{\, \tilde{G} \,}{j,m}$. For simplicity we will neglect the gauge contribution later on. For a detailed way of dealing with the trace cf. \cite{thiemann3}. 

For $l=2$ we get analogously
\begin{align}
&\Tr\left(\mathcal{D}^{(j)}(H_2(z)) \tilde{G}\right) = \sum^{j}_{m=-j}{\expval{j,m}{\, \mathcal{D}^{(j)}(H_2(z)) \tilde{G} \,}{j,m}}\notag\\
&= \sum^{j}_{m,k=-j}{e^{-i \pi m} \: d^{(j)}_{mk}(\pi - z) \: \expval{j,k}{\, \tilde{G} \,}{j,m}}
\label{eqs:27}
\end{align}

and for $l=3$
\begin{align}
&\Tr\left(\mathcal{D}^{(j)}(H_3(z)) \tilde{G}\right) = \sum^{j}_{m=-j}{\expval{j,m}{\, \mathcal{D}^{(j)}(H_3(z)) \tilde{G} \,}{j,m}}\notag\\
&= \sum^{j}_{m,k=-j}{e^{-i z m} \: \expval{j,k}{\, \tilde{G} \,}{j,m}}\,.
\label{eqs:28}
\end{align}

We will now use the large volume approximation, i.e. for \, $\I(z) \gg 1$ \, the term with $m = j$ dominates.
\begin{equation}
\Tr\left(\mathcal{D}^{(j)}(H_3(z)) \tilde{G}\right) \approx \sum^{j}_{k=-j}{e^{-i z j} \: \expval{j,k}{\, \tilde{G} \,}{j,j}}\,.\label{eqs:30}
\end{equation}

Now, how do we treat the links $l=1$ and $l=2\,$? One can argue, that due to the highly symmetric setting we should also use $m=j$ for those cases. If we do so we can make use of the following asymptotic relation \cite{angular}
\begin{equation}
d^{(j)}_{jm}(\beta) = \, (-1)^{j-m} \sqrt{\frac{(2j)!}{(j+m)! \, (j-m)!}} \: \left[\cos(\beta/2)\right]^{j+m} \left[\sin(\beta/2)\right]^{j-m}\,.
\label{eqs:31}
\end{equation}

So lets start with $l=1$. For $m=j$ we get
\begin{equation}
\Tr\left(\mathcal{D}^{(j)}(H_1(z)) \tilde{G}\right) \approx \sum^{j}_{k=-j}{e^{-i (j+k) \frac{\pi}{2}} \: d^{(j)}_{jk}(\pi - z) \: \expval{j,k}{\, \tilde{G} \,}{j,j}}\,.
\label{eqs:32}
\end{equation}

By making use of 
\begin{equation}
d^{(j)}_{mk}(\pi - z) = (-1)^{m-j} d^{(j)}_{m(-k)}(z)
\label{eqs:33}
\end{equation}

and (\ref{eqs:31}) we can analyse $d^{(j)}_{jk}(\pi - z)$ and get
\begin{align}
d^{(j)}_{jk}(\pi - z) =& \, (-1)^{j-j} \, d^{(j)}_{j(-k)}(z) = d^{(j)}_{j(-k)}(z)\label{eqs:34}\\[0.5\baselineskip]
=& \, (-1)^{j+k} \sqrt{\frac{(2j)!}{(j-k)! \, (j+k)!}} \: \left[\cos(z/2)\right]^{j-k} \left[\sin(z/2)\right]^{j+k}\notag
\end{align}

For the trigonometric functions we get for $\I(z) \gg 2$
\begin{equation}
\cos(z/2) \approx \frac{1}{2} e^{-i z/2} \quad , \quad \sin(z/2) \approx \frac{i}{2} e^{-i z/2}\,,
\label{eqs:35}
\end{equation}

which gives us the following expression
\begin{align}
\left[\cos(z/2)\right]^{j-k} \left[\sin(z/2)\right]^{j+k} &\approx \left(\frac{1}{2}\right)^{j-k} \left(\frac{1}{2}\right)^{j+k} (i)^{j+k} e^{-i \frac{z}{2} (j-k)} e^{-i \frac{z}{2} (j+k)}\notag\\[0.5\baselineskip]
&= \left(\frac{1}{2}\right)^{2j} (i)^{j+k} \, e^{-i z j}\,.
\label{eqs:36}
\end{align}

We can use this to obtain
\begin{align}
&\Tr\left(\mathcal{D}^{(j)}(H_1(z)) \tilde{G}\right) \approx \left(\frac{1}{2}\right)^{2j} \, e^{-i (z + \frac{\pi}{2}) j} \times \notag\\[0.5\baselineskip]
&\sum^{j}_{k=-j}{e^{-i k \frac{\pi}{2}} \: (-1)^{j+k} \: i^{j+k} \,  \sqrt{\frac{(2j)!}{(j-k)! \, (j+k)!}} \: \expval{j,k}{\, \tilde{G} \,}{j,j}}\,.
\label{eqs:37}
\end{align}

If we take the $e^{-i \frac{\pi}{2} j}$ term inside the sum we get $e^{-i \frac{\pi}{2} (k+j)} \: (-1)^{j+k} \: i^{j+k}$, which is equal to $(-1)^{2j+2k} \: i^{2j+2k}$, so we get
\begin{equation}
\Tr\left(\mathcal{D}^{(j)}(H_1(z)) \tilde{G}\right) \approx \left(\frac{1}{2}\right)^{2j} \, e^{-i z j} \: \sum^{j}_{k=-j}{(-1)^{2j+2k} \: i^{2j+2k} \, \sqrt{\frac{(2j)!}{(j-k)! \, (j+k)!}} \: \expval{j,k}{\, \tilde{G} \,}{j,j}}\,.
\label{eqs:38}
\end{equation}

Now let us make a few comments about the last expression. First, notice that we can reproduce the factor $e^{-i z j}$, which appears also in the original work \cite{towards} and is crucial for the derivation of the transition amplitude. The second point is that we can proceed by approximating also the second part, namely the sum over $k$, as $k = j$, which simplifies the result and the important thing is, that by doing this approximation we are not doing worse then the projection onto $m=j$ in the original work.

The calculation for $l=2$ is identical to the case $l=1$. If we now use the following two relations 
\begin{equation}
\left(\frac{1}{2}\right)^{2j} = e^{\ln(1/4) \, j} \quad, \quad(-1)^{j+k} \: i^{j+k} = (-i)^{j+k}\,.
\label{eqs:39}
\end{equation}

and apply the approximation $k=j$ we get (neglecting factors like $(-1)^{2j}$)
\begin{align}
\Tr\left(\mathcal{D}^{(j)}(H_1(z)) \tilde{G}\right) &\approx e^{-i z j + \ln(\frac{1}{4}) \, j} \, \expval{j,j}{\, \tilde{G} \,}{j,j}\,,\notag\\[0.5\baselineskip]
\Tr\left(\mathcal{D}^{(j)}(H_2(z)) \tilde{G}\right) &\approx e^{-i z j + \ln(\frac{1}{4}) \, j} \, \expval{j,j}{\, \tilde{G} \,}{j,j}\,,\label{eqs:40}\\[0.5\baselineskip]
\Tr\left(\mathcal{D}^{(j)}(H_3(z)) \tilde{G}\right) &\approx e^{-izj} \, \expval{j,j}{\, \tilde{G} \,}{j,j}\,.\notag
\end{align}

Now, let us compare these results with the calculations of the original paper \cite{towards}. There the authors used a projection onto the highest spin state $m=j$ and got a factor $\exp(-izj)$ for all links. Our calculation gives the same result using a simpler graph. Furthermore, due to our explicit calculation we see in a more precise way where this factor comes from. Now, the term $\expval{j,j}{\, \tilde{G} \,}{j,j}$ can be neglected in our case because $s(l) = t(l)$ and thus
\begin{align}
\expval{j,j}{\, \tilde{G} \,}{j,j} &= \expval{j,j}{\,  Y^{\dagger} \mathcal{D}^{(\gamma j,j)}(G_{s(l)} G^{-1}_{t(l)}) Y  \,}{j,j}\notag\\
&= \expval{j,j}{\,  Y^{\dagger} \mathcal{D}^{(\gamma j,j)}(\mathbb{I}) Y  \,}{j,j}\notag\\
&= \expval{(\gamma j,j),j,j}{\, \mathbb{I} \,}{(\gamma j,j),j,j} = 1\,,
\label{eqnew:2013}
\end{align}

where we have used the projection property of the unitary $Y$-map, cf. \cite{towards}. In \cite{towards} the authors replaced the whole gauge contribution ($\expval{j,j}{\, \tilde{G} \,}{j,j}$) from all four links by a factor $\frac{N_0}{j^3}$ based on a calculation done in \cite{newvertex}.

So far we have assumed that all our $z$-labels are the same for each link, which is a consequence of the isotropic configuration we are considering. If we investigate the anisotropic case these labels are going to be different $z=z_l$.

\subsection{Results}
Inserting the results for the traces given by (\ref{eqs:40}) and (\ref{eqnew:2013}) into the transition amplitude (\ref{eqs:25}) we get
\begin{equation}
W(z) = \left(\sum_j{(2j+1) \, e^{-2 t \hbar j (j+1) -i z j}}\right)^3\,.
\label{eqs:41}
\end{equation}

Furthermore, we have assumed that $\I(z)=p \gg \ln(1/4) \approx -1.38$, which is justified since $p > 0$, and thus all three links give the same contribution.

Now we can either apply a gaussian approximation as was done in \cite{towards}, we can investigate the amplitude numerically or we calculate (\ref{eqs:41}) explicitly using the Cauchy product, all of which yield the same result. The gaussian approximation gives the result
\begin{equation}
W(z) = (2j_0+1)^3 \, \left(\sqrt{\frac{\pi}{2 t \hbar}}\right)^3 \, e^{\frac{3 (2 t \hbar + i z)^2}{8 t \hbar}}\,,
\label{eqs:42}
\end{equation}

where $j_0$ is given by
\begin{equation}
j_0 = - \frac{1}{2} + \frac{\text{Im}(z)}{4 t \hbar}\,.
\label{eqs:43}
\end{equation}

In order to get meaningful results we now have to normalize the amplitude. For this we use the following expression for the norm of a heat kernel coherent state given on a single link \cite{thiemann3}
\begin{equation}
\left\|\psi^{\tilde{t}}_g\right\| = \frac{4 \sqrt{\pi} e^{\tilde{t}/4}}{\tilde{t}^{3/2}} \frac{1}{\sinh(\tilde{p})} \frac{\tilde{p}}{2} e^{\frac{\tilde{p}^2}{\tilde{t}}}\,,
\label{eqs:44}
\end{equation}

where we have taken just the leading order term with $n=0$, cf. \cite{thiemann3}. The small $g$ in the above formula corresponds to our $SL(2,\mathbb{C})$ element $H_l(z)$ and the heat kernel time $\tilde{t}$ is related to our $t$ via $\tilde{t} = 2 t \hbar$. A detailed analysis reveals furthermore that the $\tilde{p}$ in (\ref{eqs:44}) corresponds our $\frac{p}{2}$.  

We will start with a numerical analysis of (\ref{eqs:41}). Therefore, we plot $\mathcal{A}(z)$, which we define as the absolute value of $W(z)$ divided by the norm to the third, because our graph has three links. 
\begin{equation}
\mathcal{A}(z) = \frac{\abs{W(z)}}{\left\|\psi^{\tilde{t}}_g\right\|^3}\,.
\label{eqs:45}
\end{equation}

\begin{figure}[ht]
\includegraphics[width=0.45\textwidth]{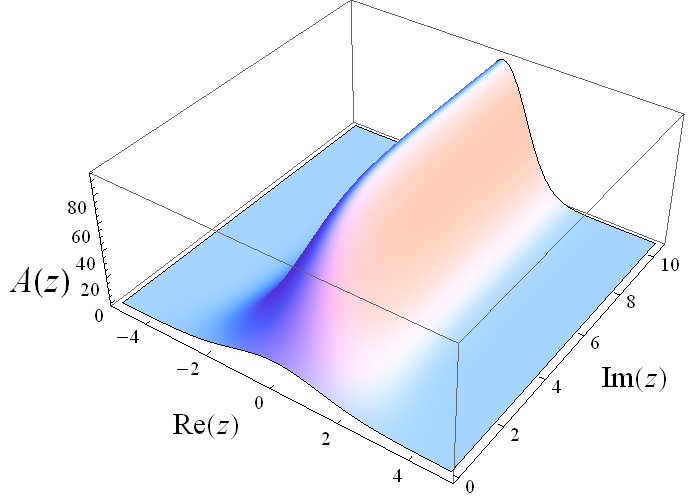}
\caption{Normalized amplitude $\mathcal{A}(z)$ for the Cube with $j_{max} = 150$ and $t=1$}
\label{fig2}
\end{figure}

\begin{figure}[ht]
\includegraphics[width=0.45\textwidth]{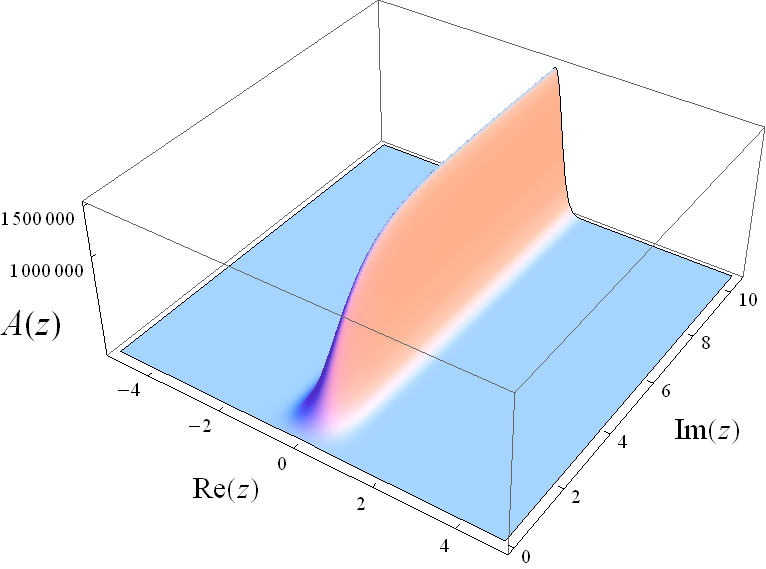}
\caption{Normalized amplitude $\mathcal{A}(z)$ for the Cube with $j_{max} = 150$ and $t=0.1$}
\label{fig3}
\end{figure}

We set $\hbar = \frac{1}{2}$ and truncate the sum (\ref{eqs:41}) at $j_{max} = 150$ where one has to make sure that $j_0 < j_{max}$, (\ref{eqs:43}) holds. If we set the heatkernel time $t=1$ we get FIG.(\ref{fig2})
and with a heatkernel time $t=0.1$ we see that the peak becomes sharper FIG.(\ref{fig3}).

Now, recall that $\R(z)$ corresponds to the extrinsic curvature of our model and thus we find $\R(z) \propto \dot{a} = 0$ for all volumes $\I(z) \propto a^2$. Hence, we find that our universe is static. As we would expect. We don't have any matter or a cosmological constant, nor anisotropies.

What is remarkable now is the decrease of the amplitude towards small $\I(z)$. To see this more clearly insert (\ref{eqs:43}) into (\ref{eqs:42}) and calculate $\mathcal{A}(z)$ using (\ref{eqs:44}). The result is
\begin{equation}
\mathcal{A}(c,p) = \frac{1}{4} \sinh^3\left(\frac{p}{2}\right) e^{-\frac{3p}{2}} \, e^{-\frac{3 c^2}{8 t \hbar}}\,.
\label{eqs:46}
\end{equation}

\begin{figure}[ht]
\includegraphics[width=0.45\textwidth]{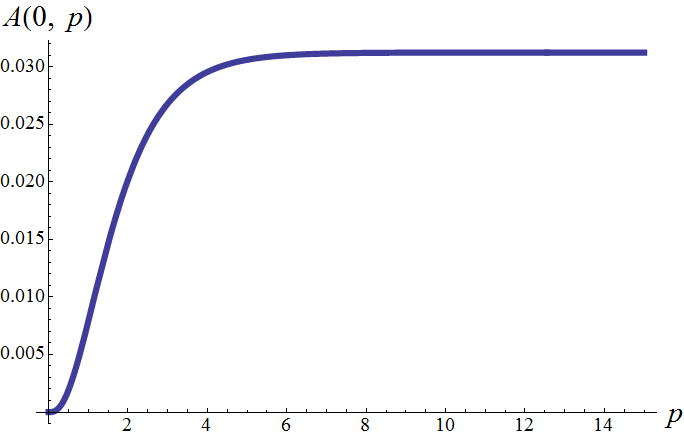}
\caption{The transition amplitude as a function of scale factor, for a typical fixed $c$, here chosen to be zero. We find the amplitude is not supported on $p=0$, indicating that there can be no transition to singularity.}
\label{fig4}
\end{figure}

Plotting $\mathcal{A}(0,p)$ we get the shape in the $p$-direction which shows us the drop off for small scale factors FIG.(\ref{fig4}).

What does this tell us about the quantum dynamics of our model? Recall the definition of the transition amplitude between two quantum states of geometry $\Psi_i$ and $\Psi_f$, (\ref{eqs:1}). One finds that the main contributions come from those configurations corresponding to classical geometries, namely $\dot{a} = 0$ and large scale factors $a$. The remarkable result is now, that the transition amplitude decreases for small scale factors and has zero support on $a=0$. This is a statement about the occurrence of singularities in our model in that it tells us that a transition to a singularity is ruled out dynamically.

The transition amplitude now gives us the possibilities for quantum fluctuations from one scale factor to another one. It doesn't give us the dynamical evolution of our universe. The classical notion of dynamics, i.e. $\dot{a}$, is encoded in the phase space coordinates. In the same sense as a transition amplitude in QFT does not give us a temporal information of the time evolution of a certain process.

Certainly this result has to be strengthened by future investigations, where it remains to show that one can circumvent the large spin approximation. In fact, one can calculate the traces in (\ref{eqs:25}) in our model explicitly, thanks to the fact that our graph has just one node and thus $\expval{j,k}{\, \tilde{G} \,}{j,m} = \expval{j,k}{\,  Y^{\dagger} \mathcal{D}^{(\gamma j,j)}(G_{s(l)} G^{-1}_{t(l)}) Y  \,}{j,m} = \expval{(\gamma j,j),j,k}{\, \mathcal{D}^{(\gamma j,j)}(\mathbb{I}) \,}{(\gamma j,j),j,m} = \delta_{km}$ holds. This way one can in principle avoid the large spin approximation and indeed finds that the amplitude is still peaked on $c = \R(z) = 0$. However, the shape along the $p = \I(z)$ direction changes, a problem which probably has to be solved by the use of a different normalization.

\section{Discussion}
We showed in this paper that within the spinfoam cosmology approach the Daisy graph is sufficient to reproduce the vacuum Friedmann equation for a flat 3-space in the isotropic setting and thus may also be useful for the investigation of anisotropic models. The Daisy graph is perfectly suited to the relaxation of the restriction to isotropic models just by using three different holomorphic labels $z_i$ at each link. This will be our setting in the following paper \cite{ournextpaper}. Furthermore, we showed how one can reproduce the missing curvature term in the curved model, described by the Dipole graph, without using higher terms in the spin expansion. The right dependence of the real and imaginary part of the holomorphic labels $z$ will also be important for the description of the anisotropic model. It has been suggested that the curvature term arises as a `higher order' effect in the prior models. However, this derivation applies even in the flat case, which would contradict the agreement with classical dynamics at large scale factor. 

The main result of this work, however, is the statement about the avoidance of singularities in our model. The dynamics of the Daisy graph show zero support for a transition amplitude from a finite scale factor to zero, therefore the singularity itself is not accessed by dynamics. Since the model under consideration does not include matter terms, there can be no direct comparison made with the bouncing models of LQC. It remains to be seen at this stage whether this result is an artefact of approximations made, or is a deeper feature of the full covariant dynamics. A singularity resolution theorem or results analogous to those of \cite{ParamLQCNeverSingular} would be a strong achievement for the theory. Since the interpretation of a transition amplitude between an in and out state in this model is asymptotic, and we only consider the first order in perturbation theory (in terms of graph expansion, vertex expansion etc) one cannot make any strong claim of singularity avoidance. However, at this order there is a hint of a resolution in the manner of a bounce: Consider a sequence of transitions between scale factors each described with a single transition amplitude. This will be a random walk in the space of scale factors with transition probabilities as described by the distribution in FIG.(\ref{fig3}). As the scale factor tends to zero, the probability of moving to a larger scale factor increases, being unity in the limit of zero scale factor. Thus we see that the probability of a collapsing universe continuing to collapse tends to zero, and the probability of expansion tends to one, as we approach the classical singularity, and thus the universe will undergo a bounce. As we have noted, this is a preliminary result, and may not survive extension of the model beyond the simple expansion used here, but nonetheless is encouraging in its similarity to the singularity resolution seen in LQC. Obviously this isn't the complete picture, as we are moving within points of configuration space at which $\dot{a}=0$ without giving the detailed dynamics between. This behaviour in which the zero volume state is excluded from solutions hints that singularity resolution may well be a feature of these models, as has recently been argued in \cite{CarloFrancescaRecent}.  

Needless to say, that there are numerous open questions that need to be investigated within the spinfoam cosmology approach, such as the treatment of higher orders in the vertex expansion and of course the inclusion of further spinfoam histories. It is hoped that this will shed some light on the connection with the (quantum reduced) canonical approach as put forward by \cite{quantred, quantred2} and also the GFT cosmology in \cite{gft}. In the long term perspective the coupling of matter has also high priority if one aims to seriously do quantum cosmology using such models.

\section*{Acknowledgments}
The authors would like to thank Carlo Rovelli and Francesca Vidotto for useful comments and discussion. DS gratefully acknowledges support from a Templeton Foundation grant. The authors are indebted to the referees whose extensive input has greatly improved this paper.

\begin{appendix}
\section{Details of the calculations}
In this section we present the calculation of the $SL(2,\mathbb{C})$ elements $H_l(z)$, (\ref{eqs:3}). As explained in section 2.1 the normal vectors $n_l$ and $\tilde{n}_l$ are obtained via a $SO(3)$ transformation of $\hat{e}_z$ and the $SU(2)$ elements $u_l$ and $\tilde{u}_l$ are related to these $SO(3)$ transformations. We start with $\hat{e}_z \mapsto \hat{e}_x$ and $\hat{e}_z \mapsto -\hat{e}_x$.
\begin{equation}
\begin{pmatrix} 1 \\ 0 \\ 0 \end{pmatrix} = R^{(x)}_y \hat{e}_z = \begin{pmatrix}\cos(\phi) & 0 & \sin(\phi) \\0 & 1 & 0 \\-\sin(\phi) & 0 & \cos(\phi)\end{pmatrix} \cdot \begin{pmatrix}0 \\ 0 \\ 1\end{pmatrix} \, \stackrel{\phi = \frac{\pi}{2}}{=} \, \begin{pmatrix}0 & 0 & 1 \\0 & 1 & 0 \\-1 & 0 & 0 \end{pmatrix} \cdot \begin{pmatrix}0\\0\\1\end{pmatrix}\label{eqa:4}
\end{equation}

\begin{equation}
\begin{pmatrix}-1\\0\\0\end{pmatrix} = R^{(-x)}_y \hat{e}_z = \begin{pmatrix}\cos(\phi) & 0 & \sin(\phi)\\0 & 1 & 0\\-\sin(\phi) & 0 & \cos(\phi)
\end{pmatrix} \cdot \begin{pmatrix}0\\0\\1\end{pmatrix} \, \stackrel{\phi = \frac{\pi}{2}}{=} \, \begin{pmatrix}0 & 0 & -1\\0 & 1 & 0\\1 & 0 & 0\end{pmatrix} \cdot \begin{pmatrix}0\\0\\1
\end{pmatrix}\label{eqa:5}
\end{equation}

We calculate the two corresponding $SU(2)$ elements for the $SO(3)$ rotation matrix $R$ with the formula \cite{carmeli, urbantke}
\begin{equation}
u = \mp \, \frac{\left(\mathbb{I}_2 + \sigma^r \sigma^s R_{rs}\right)}{\left(2 \sqrt{1 + \Tr R}\right)} \: \in \: SU(2)\,.
\label{eqa:6}
\end{equation}

We get for $R^{(x)}_y$ and $R^{(-x)}_y$
\begin{equation}
R^{(x)}_y \: \rightsquigarrow \: u^{(x)} = \mp \, \frac{1}{\sqrt{2}} \begin{pmatrix}
	1 & -1 \\
1 & 1
\end{pmatrix} \quad , \quad
R^{(-x)}_y \: \rightsquigarrow \: u^{(-x)} = \mp \, \frac{1}{\sqrt{2}} \begin{pmatrix}
	1 & 1 \\
-1 & 1
\end{pmatrix}\,.
\label{eqa:8}
\end{equation}

Analogously one calculates $\hat{e}_z \mapsto \hat{e}_y$ and $\hat{e}_z \mapsto -\hat{e}_y$ with the resulting $SU(2)$ elements
\begin{equation}
R^{(y)}_x \: \rightsquigarrow \: u^{(y)} = \mp \, \frac{1}{\sqrt{2}} \begin{pmatrix}
	1 & i \\
i & 1
\end{pmatrix} \quad, \quad
R^{(-y)}_x \: \rightsquigarrow \: u^{(-y)} = \mp \, \frac{1}{\sqrt{2}} \begin{pmatrix}
	1 & -i \\
-i & 1
\end{pmatrix}\,.
\label{eqa:10}
\end{equation}

Finally, we have to calculate $\hat{e}_z \mapsto \hat{e}_z$ and $\hat{e}_z \mapsto -\hat{e}_z$ but it is clear that $R^{(z)}_y$ is given by
\begin{equation}
R^{(z)}_y = \mathbb{I}_3
\label{eqa:11}
\end{equation}

and thus the corresponding $SU(2)$ elements is 
\begin{equation}
u^{(z)} = \mp \mathbb{I}_2\,.
\label{eqa:12}
\end{equation}

Now we have to connect each two $SU(2)$ elements with one link. We connect those vectors who are co-linear. Furthermore, we need
\begin{equation}
e^{-i z \frac{\sigma^3}{2}} =\begin{pmatrix}
	e^{-i \frac{z}{2}} & 0 \\
0 & e^{i \frac{z}{2}}
\end{pmatrix}\,.
\label{eqa:13}
\end{equation}

The inverse matrices are given by
\begin{equation}
\left(u^{(-x)}\right)^{-1} = \mp \, \frac{1}{\sqrt{2}} \begin{pmatrix}
	1 & -1 \\
1 & 1
\end{pmatrix} \quad , \quad
\left(u^{(-y)}\right)^{-1} = \mp \, \frac{1}{\sqrt{2}} \begin{pmatrix}
	1 & i \\
i & 1
\end{pmatrix} \quad , \quad
\left(u^{(-z)}\right)^{-1} = \mp \mathbb{I}_2\,,
\label{eqa:16}
\end{equation}

where we have chosen (\ref{eqa:16}) corresponding to the inverse elements of $u^{(x)}$ and $u^{(y)}$ resp. The reason is that formula (\ref{eqa:6}) is not valid for a rotation of $\pi$ about the $x-$ or $y-$axis.

We get for the $SL(2,\mathbb{C})$ elements
\begin{align}
H_1(z) &= u_1 \, e^{-i z \frac{\sigma^3}{2}} \, \tilde{u}^{-1}_1 = u^{(x)} \, e^{-i z \frac{\sigma^3}{2}} \,  \left(u^{(-x)}\right)^{-1}\notag\\[0.5\baselineskip]
&= \frac{1}{2} \begin{pmatrix}
	1 & -1 \\
1 & 1
\end{pmatrix} \cdot \begin{pmatrix}
	e^{-i \frac{z}{2}} & 0 \\
0 & e^{i \frac{z}{2}}
\end{pmatrix} \cdot \begin{pmatrix}
	1 & -1 \\
1 & 1
\end{pmatrix}\notag\\[0.5\baselineskip]
&= \frac{1}{2} \begin{pmatrix}
	e^{-i \frac{z}{2}} - e^{i \frac{z}{2}} & -e^{-i \frac{z}{2}} - e^{i \frac{z}{2}} \\
e^{-i \frac{z}{2}} + e^{i \frac{z}{2}} & -e^{-i \frac{z}{2}} + e^{i \frac{z}{2}}
\end{pmatrix}\notag\\[0.5\baselineskip]
&= \begin{pmatrix}
	- i \sin\left(\frac{z}{2}\right) & -\cos\left(\frac{z}{2}\right) \\
\cos\left(\frac{z}{2}\right) & i \sin\left(\frac{z}{2}\right)
\end{pmatrix}\notag\\[0.5\baselineskip]
&= -i \left(\sin\left(\frac{z}{2}\right) \sigma^3 + \cos\left(\frac{z}{2}\right) \sigma^2\right)
\label{eqa:17}
\end{align}

and for $l=2$ and $l=3$ we get analogously
\begin{align}
H_2(z) &= u_2 \, e^{-i z \frac{\sigma^3}{2}} \, \tilde{u}^{-1}_2 = u^{(y)} \, e^{-i z \frac{\sigma^3}{2}} \,  \left(u^{(-y)}\right)^{-1}\notag\\[0.5\baselineskip]
&= i \left(\cos\left(\frac{z}{2}\right) \sigma^1 - \sin\left(\frac{z}{2}\right) \sigma^3\right)\,,
\label{eqa:18}
\end{align}

\begin{align}
H_3(z) &= u_3 \, e^{-i z \frac{\sigma^3}{2}} \, \tilde{u}^{-1}_3 = u^{(z)} \, e^{-i z \frac{\sigma^3}{2}} \,  \left(u^{(-z)}\right)^{-1}\notag\\[0.5\baselineskip]
&= \begin{pmatrix}
	e^{-i \frac{z}{2}} & 0 \\
0 & e^{i \frac{z}{2}}
\end{pmatrix}\,.
\label{eqa:19}
\end{align}

\end{appendix}

\bibliography{references}
\bibliographystyle{ieeetr}

\end{document}